# UMG silicon for solar PV: from defects detection to PV module degradation

*Eduardo Forniés[1]\*, Carlos del Cañizo[2], Laura Méndez,[1] Alejandro Souto[3], Antonio Pérez Vázquez[3], Daniel Garrain[4]*

[1]*Aurinka PV Group, Marie Curie 19, Rivas-Vaciamadrid (Madrid), Spain*
[2]*Instituto de Energía Solar – Universidad Politécnica de Madrid, Avda. Complutense, 30, Madrid, Spain*
[3]*Ferroglobe, Arteixo-La Coruña, Spain*
[4]*CIEMAT, Av.de la Complutense 40, Madrid, Spain*
*Corresponding author: \*email: efornies@aurinkapv.com*

Upgraded metallurgical grade silicon (UMG-Si) for photovoltaic (PV) solar applications has been manufactured through the metallurgical route by means of the process developed by Ferrosolar. In an ambitious mass production test, performed in commercial solar cells and modules production lines, the silicon was proven to be appropriate for photovoltaics applications (Forniés et al., 2019 Mass production test of solar cells and modules made of 100% umg silicon. 20.76% record efficiency. Energies 12), reaching, in a conventional production line, up to 20.76% of solar cell efficiency with multicrystalline cells made of 100% UMG silicon. In this paper we present more results from the mentioned massive test. Defect engineering is being applied to improve the bulk lifetime of the UMG wafers and to guide in the identification of the limiting defects in the material. Moreover, the modules produced with 100% UMG silicon solar cells were installed together with the modules produced in the same production line with polysilicon material to assess the degradation of the UMG silicon when compared to polysilicon. After 24 months of outdoor PV generation, the degradation, in terms of Performance Ratio at 25ºC ($^{25}$PR) diminution, has been the same for both types of modules. Additionally, a Life Cycle Assessment (LCA) has been performed for this UMG silicon and state-of-the-art Siemens polysilicon to compare the environmental impact of both silicon feedstocks. The results presented in this paper; chemical analysis of wafers, defect engineering, low degradation, average efficiency and environmental assessment, lead to a complete study of UMG silicon, confirming its potential to be used as raw material for PV applications.

## 1. Introduction

No argument arises when stating the important role of Photovoltaics (PV) in the strategy to transform the energy generation systems. For instance, solar has dominated the overall energy addition in the power sector in the last years and the forecast of future additions expects a total installed capacity of 8.5 TW of PV in 2050, representing more than 40% of the total installed power capacity (including fossil and nuclear) (IRENA, 2019). A more detailed study (Breyer et al., 2018), in a hypothetic primary electricity generation scenario of 100% renewable energy, forecasts a 69% global share of PV by 2050, which is by far, the most rapid growth of a electricity generation technology. To achieve this target, it is essential to keep on lowering the Levelized Costs of Electricity (LCoE) of PV, that has already demonstrated to be even lower than most of the electricity generation alternatives, either renewables or fossil (Kost et al., 2018). The increasing efficiency of solar cells like PERC (passivated emitter and rear cell (Altermatt et al., 2018)), SHJ (heterojunction), and TOPCon (tunnel oxide passivated contact (Zeng et al., 2017)) and modules (half cells, bifacial (Chen et al., 2019), etc.) is contributing to further decreasing of LCoE. The LCoE reduction can be also achieved by reducing the production costs of materials, in particular that of silicon, as it accounts nowadays for approximately 20% of the crystalline silicon module cost. For more than 15 years, FerroSolar has developed its own metallurgical route of silicon purification, resulting in a material that is suitable for PV applications. It must be stated that this silicon, with a purity of 99.9999%, has a higher concentration of dopants (boron and phosphorous) and metals than conventional polysilicon, which contributes to a lower minority carrier bulk lifetime ($\tau_{min,b}$) of the multicrystalline wafer grown with this feedstock. Without any defect engineering process, the lower $\tau_{min,b}$ is translated into a lower efficiency at the cell level. Thus, to counterbalance the slight performance reduction compared with polysilicon, the UMG-Si should offer other advantages; lower productions costs, lower capital expenditure and lower global greenhouse gases (GHG) emissions. Besides, no other detrimental effects related to UMG-Si should appear during outdoors energy production, like additional power or light at elevated temperature induced degradation, (PID (Luo et al., 2017), LeTID (Jensen et al., 2017; Kersten et al., 2015)).

Some results have been demonstrated in previous papers (Forniés et al., 2018; Forniés et al., 2019). The measured average efficiency of UMG cells compared with polysilicon cells was 18.40 vs 18.49% respectively for aluminium back surface field (Al- BSF) cells and 20.13 vs 20.41 % for PERC technology. As it has been mentioned in those publications, these results were obtained without any adjustment of production parameters in the production lines where the solar cells were manufactured. It is obvious that different concentration in dopants and metals requires an adjustment of parameters, especially for thermal processes (P diffusion, annealing, cofiring), to obtain the higher potential performance of the UMG-Si solar cells. Thus, considering that the cells were made 100% with UMG-Si and no production line adjustment was carried out, the achieved reduction of power can be regarded as very low.

Aurinka, in collaboration with IES-UPM, NTC-UPV and GÜNAM-METU is currently involved in a research project aimed to develop the solar cell process tailored to multi wafers made of UMG-Si. Black silicon, extended phosphorous diffusion gettering and current induced recovery are some of the techniques to be studied in order to get the most out of the cell performance.

In this paper, the authors carry out a deeper research on the contaminants present in this silicon and how they are influencing the bulk lifetime. Firstly, two different types of wafers were manufactured: type A wafers were manufactured with the standard UMG silicon of Ferrosolar, by a process that is more extensively described in present work. Type B wafers were produced in the same process, butdeliberately introducing more contaminated silicon material. The purpose of manufacturing contaminated wafers, mainly with metals, is to give a range for solar cell performance depending on the concentration of metals and dopants in the silicon feedstock. This would be the starting point to obtain an empirical correlation between the solar cell efficiency and the contaminant concentration. Chemical analysis was performed on the two types of wafers to determine mainly the concentration of dopants and metals. Secondly, Injection-Dependent Lifetime Spectroscopy (IDLS) was used to elucidate the possible contaminants present in the wafers. Then, both types of wafers, type A and B, were sent to a manufacturer to produce Al-BSF solar cells. The electrical results of both types of solar cells are presented in this study.

Subsequently, solar modules were manufactured with type A solar cells (named type A modules) and their performance compared to the same type of modules but with solar cells made of polysilicon, as published elsewhere (Forniés et al., 2018). The two types of modules have been installed together in the same PV installation and the results of degradation after 24 months of operation are presented also in this study. It is known that different types of degradations affect the modules once they are installed outdoors. The authors want to focus on light induced degradation, LID (Sopori et al., 2012), light and elevated temperature degradation LeTID or hydrogen induced degradation, HID (Jensen et al., 2017; Ciesla et al., 2018) and power induced degradation PID (Luo et al., 2017), all of them considered as early loss degradations that happen before 1 year of outdoor exposure and that can be affected by the contaminants in the silicon material.

To finalize, we present a Life Cycle Assessment of the UMG-Si production process of Ferrosolar and compare it with the most extended process of polysilicon production (closed-loop Siemens process with hydrochlorination), focusing on the results for Climate Change.

## 2. Methodology

The present study has been carried out on UMG silicon feedstock. The process of silicon purification is explained roughly in section 2.1. In Fig. 1 the procedure of the whole study is depicted.

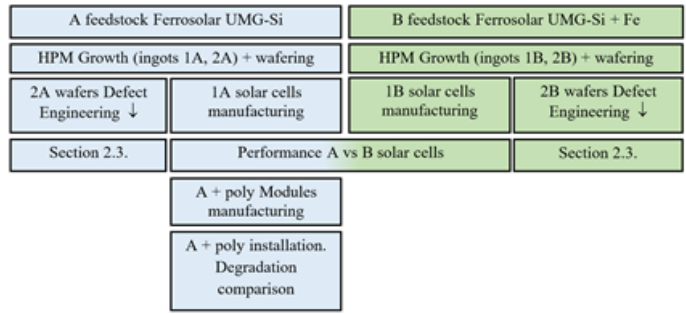

*Figure 1. Overview of the procedure followed in this study. HPM stands for High Performance Multicrystalline. The starting point is the silicon purification explained in section 2.1. A feedstock: silicon from Ferrosolar´s standard process, B feedstock: same silicon but intentionally mixed with contaminated silicon in the charge of multi growth.*

2.1. Silicon purification

Two different silicon feedstocks were prepared for this test: A feedstock and B feedstock. A feedstock is obtained by the Ferrosolar´s standard silicon purification process, described in this section. B feedstock is obtained by mixing the A silicon with contaminated silicon.

Ferrosolar´s process for silicon purification from metallurgical grade (MG-Si) to solar grade (SoG-Si) is a sequence of successive steps to achieve the purification degree required (Ceccaroli et al., 2016; Hoffmann et al., 2015). Chunks and ingots of silicon of adequate purity for the photovoltaic industry are the final result of this process.

Ferrosolar process begins with the crushing of MG Si, to adapt its grain size to subsequent treatments. Crushers have been carefully designed to minimize contamination. The final grain size is 100% less than 1 cm.

Next, this silicon is charged in induction furnaces with a capacity up to one ton of silicon. Ferrosolar uses reactive slags in graphite crucibles with the objective of a selective removal of certain impurities, including boron. Once the treatment is finished, the molten silicon is transferred to a suitable vessel and directionally solidified. The degree of purification in this last step depends on its segregation coefficient of a given impurity. The overall result of this first stage of slagging and controlled solidification is the reduction of boron and most of the metallic impurities (Al, Ca, Fe, Ti, etc), as well as the partial reduction of other impurities like phosphorus (50% reduction) among others.

In the next step, the directionally solidified silicon is grinded bellow 1.5 cm and leached in acid, washed in deionized water and dried to prepare it for next stage which is a thermal treatment at temperatures in the range 1550°C – 1700°C under high vacuum (Souto et al. 2014). The main objective of this process is the removal of phosphorus, given its high volatility under vacuum at high temperatures. Other impurities such as Al, Ca, K, and Na, are also partially evaporated. In this step, the silicon is firstly introduced into a sealed chamber equipped with internal atmosphere control. This chamber, acts as the silicon charging system and is connected to the high vacuum furnace through a valve. Once the silicon has been loaded, the chamber is purged to reduce its pressure before opening the connection valve with the vacuum furnace. When the purge has been completed, the internal pressure in the charge chamber is adjusted to 3000 Pa to match that of the vacuum furnace. When internal pressures on both sides are equivalent the valve opens and the silicon is transferred into the vacuum furnace, where is placed in a high-density graphite crucible with a capacity up to 1000 kg of silicon.

When the transfer of silicon has been completed the valve closes, and the vacuum furnace is ready to start with the thermal treatment. The furnace is equipped with two independent systems to control temperature; a W-Re thermocouple placed close to the crucible, and an optical pyrometer that measures the temperature through a quartz crystal window. Internal pressure is initially 3000 Pa, and this pressure is kept constant until the melting of the silicon has been completed and the temperature stabilized at the chosen setpoint. At this point, to increase the evaporation kinetics it is necessary to lower the internal pressure in the furnace from 3000 Pa to below 5 Pa . In fact, evaporation of phosphorus reaches its maximum rate at pressures from 1 to 5 Pa at the range of temperatures of these treatments. Evaporation lasts between 4 and 8 hours, depending on the initial phosphorus concentration of the silicon.

At the end of the evaporation, the furnace is refilled with Ar to an internal pressure of 3000 Pa. Then, it is connected to a solidification furnace and the purified silicon is casted and directionally solidified. The result of this evaporation/controlled solidification process is a silicon block of 500-1000 kg depending on the furnace capacity. Finally, this block is placed in a diamond wire saw machine to remove the laterals, upper and lower parts where the residual impurities are concentrated. These contaminated parts will be recycled and re-melted in previous stages of purification. The silicon thus obtained after the cutting step has very low boron, phosphorus and metals content and is very suitable for photovoltaics.

The process described above has been used to produce the raw material for wafers A. Specifications of the silicon lot used to make the growth were:

[B]      < 0.2 ppmw
[P]      < 0.3 ppmw
$\sum$[Me]    < 0.5 ppmw

As mentioned before, silicon B, is prepared by mixing the silicon A with contaminated silicon. The contaminants are mainly Fe, Cr and Ni. A representative sample of both types of silicon (silicon A and contaminated silicon) was measured by means of ICP-MS. Then the concentration of metals of silicon B is calculated according to the blending percentage with contaminated silicon (see Table 1).

| Silicon feedstock | Fe ppm$_w$ | Al ppm$_w$ | Ti ppm$_w$ | Cu ppm$_w$ | Cr ppm$_w$ | Ni ppm$_w$ |
|---|---|---|---|---|---|---|
| B | 13.4 | 0.5 | 0.1 | 0.3 | 12.4 | 4.1 |
| A | <0.1 | 0.2 | <0.1 | <0.1 | <0.1 | <0.1 |

*Table 1. Calculated metals concentration of silicon feedstock A and B.*

2.2. High Performance Multi (HPM) wafer production

The HPM process (Buchovska et al., 2017; Forniés et al., 2018; Forniés et al., 2019) has been used to grow multicrystalline silicon ingots. One ingot, named as ingot 1A, was grown using the silicon feedstock specified in section 2.1. A second ingot (1B) was obtained by mixing the UMG silicon with silicon contaminated with metals (mainly Fe, Cr, Ni). From now, A and B will refer to the silicon feedstock, independently of talking about ingots, bricks, wafers, cells or modules.

As the silicon feedstock used for this test is a compensated silicon (similar concentration of boron than phosphorous), certain quantities of Ga (from 5 to 10 ppmw) and P (0.12 ppmw) were added to control the resistivity variation along the ingots (Forster et al., 2012a; Forster et al., 2012b). To calculate the dopant additions, Klaasen´s equations (Klaassen, 1992) for a compensated silicon have been used to model the resistivity along the ingot height, resulting in the following final dopants concentrations in the charge:

[P] = 0.42 ppm$_w$
[B] = 0.17 ppm$_w$
[Ga]=   9 ppm$_w$

Note that P was added to the charge to compensate the boron concentration, while the addition of Ga is used to compensate the accumulation of phosphorous at the end of the ingot. In Fig. 2 the two cases, with Ga and without Ga, are plotted. It is clear how Ga, due to its low segregation coefficient (k=0.008), compensates the cumulated P at the end of the ingot, contributing to enclose the resistivity values at the end of the ingot within the standard specification range (1-3 $\Omega \cdot$cm).

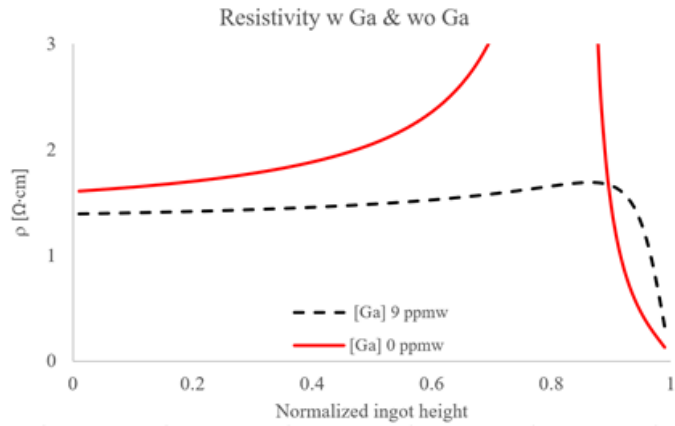

*Figure 2. Resisitivy profile applying the Klassen´s model along the ingot height for two samples; solid line : [P]=0.42ppmw, [B]=0.17ppmw, [Ga] =0ppmw, dashed line: [P]=0.42ppmw, [B]=0.17 ppmw, [Ga] =9ppmw. Standard specifications: 1-3 Ω·cm*

Some studies have questioned the accuracy of Klaasen's model for the mobilities in compensated material (Forster et al, 2012a, Fourmond et al. 2011), but that can be so for compensation levels (defined as the sum of acceptors and donors divided by their difference) higher than around 10, while in this case compensation levels are at most 5 in the top of the ingot.

After growth, the ingots were cut into 25 bricks (G5). Two bricks of each ingot (1A and 1B) were selected to perform this study. Lifetime of the bricks was measured by means of transient µ-PCD (Semilab WT-2000p) with an excitation laser source of 904 nm, (see Fig. 3). The Semilab´s lifetime mapping can be seen in the same figure, together with the average along the height of two faces for each brick. The lifetime mapping shows a lower lifetime of ingot 1B compared with ingot 1A which is in line with a higher concentration of contaminants in silicon 1B. Note that, as this is a HPM growth, the bottom fraction of the brick with low lifetime (red fraction) is thicker than it would be in a standard directional solidification growth. The reason is that the lower section of the crucible is filled with a layer of polysilicon seeds of about 20 mm (Forniés et al., 2019).

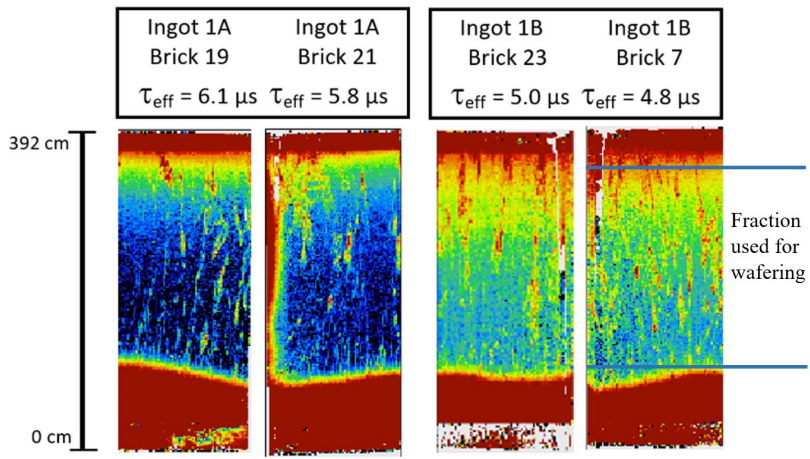

*Figure 3. Lifetime (LT) mapping by means of µW-PCD Semilab WT-2000p. Top inset express the LT averages of two faces of the same brick. As an example, the fracction used for one brick is shown in blue lines.*

The resistivity values measured by Eddy current are presented in Fig.4. As two different bricks from each ingot were used for this test, the resistivity results are the average of the bricks used. Although within specifications (1-3 Ω·cm), it is clear that bricks from ingot B, showed a lower resistivity than that measured on ingot A, which matches fairly well with the Klaassen´s model. The explanation is that the final resistivity is quite sensible to the initial concentration of boron, which can vary slightly from one furnace charge to another.

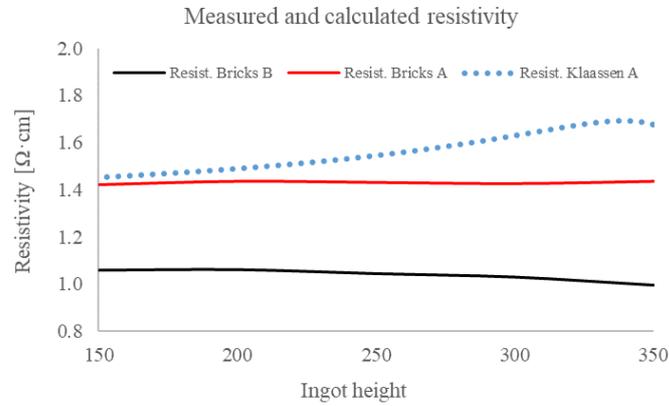

Figure 4. Average Eddy current resisitivity of bricks A and B

After analysis, the bricks were cut into wafers by means of slurry wire saw machines.

2.3. Chemical analysis and defects characterization

To perform the chemical analysis and the defect engineering tests, two additional ingots were grown with the same silicon feedstocks than in the previous section. In this case, ingots were named 2A and 2B, being the label A or B the reference to the silicon used. The brick number 19 from each ingot was taken for wafering (see figure 5). Some of the wafers from different heights were crushed to a grain size below 1 cm and chemically etched in a highly diluted solution of HF and HNO3 for surface cleaning. Then, different replicas of each sample were prepared, and dissolved in a high purity HF/$HNO_3$ mixture and dried. Finally, the dry residue resulting from the chemical digestion was re-dissolved in diluted high purity HCl.. Resulting solutions were analyzed by inductive coupling plasma mass spectroscopy (ICP-MS). Chemical analysis of metals (Al, Ca, Fe, Ti, Cu, Cr, Co) were performed by inductive coupling plasma optical emission spectroscopy (ICP-OES) in our labs, while dopant elements (B, P and Ga) were measured by ICP MS QQQ (QQQ stands for triple quadrupole) model Agilent 8800 ICP-QQQ

Table 2 shows the results for B, P and Ga. Regarding metals, their concentrations are well below quantification limits of our ICP OES (2 $ppm_w$). Further analysis of metals in Agilent 8900 ICP-QQQ revealed a high scattering of measures in different replicas of the same sample. Although the samples preparation and analysis were made in a clean room, the scattering reveals the difficulties of these types of analysis, showing that small contaminations in the sample surface, environment, tools, etc, can lead to different element concentration measurements in different replicas. Nevertheless, the results on Cr, Co and Ni (not shown in the table) presented very low concentration in all the replicas, all of them below 0.008 $ppm_w$ and also a low standard deviation, leading to the conclusion that those elements can be discarded as the cause of lifetime reduction in wafers.

|  | B | P | Ga |
|---|---|---|---|
| **Si Type A** (sample 1) | 0.23 | 0.39 | 0.15 |
| **Si Type A** (sample 2) | 0.22 | 0.35 | 0.15 |
| **Si Type A** (sample 3) | 0.23 | 0.37 | 0.14 |
| **Si Type B** (sample 1) | 0.32 | 0.29 | 0.17 |
| **Si Type B** (sample 2) | 0.30 | 0.34 | 0.18 |
| **Si Type B** (sample 3) | 0.24 | 0.36 | 0.18 |

*Table 2. Dopants chemical analysis in silicon wafers type A and B in ppm$_w$*

To overcome the limitations of UMG silicon by means of defect engineering, it is of major importance to identify the type of defects present in this silicon; grain boundaries, dislocation clusters, recombination active contaminants, etc. Moreover, it is known that although hydrogenation has beneficial effects on the passivation of some of the wafer´s defects, i.e. grain boundaries (Sio et al., 2017), it is also involved in the degradation at light and elevated temperature LeTID, also called hydrogen induced degradation (HID) (Ciesla et al., 2018) where other species, likely metals, play an important role in combination with hydrogen (Schmidt et al., 2019). An analysis based on Injection-Dependent Lifetime Spectroscopy (IDLS) (Morishige et al., 2017; Murphy et al., 2012) has been carried out to support in the identification of the limiting defects in the material, and to explore the differences between type A and type B wafers

| 1 | 2 | 3 | 4 | 5 |
|---|---|---|---|---|
| 6 | 7 | 8 | 9 | 10 |
| 11 | 12 | 13 | 14 | 15 |
| 16 | 17 | 18 | **19** | 20 |
| 21 | 22 | 23 | 24 | 25 |

*Figure 5. Sketch of the cutting of an ingot. Red bold line shows the face where the LT average has been calculated*

The ingots disposition is depicted in Fig.5, where the faces used to measure the µW-PCD lifetime are marked in red. The average of the three faces together with the average resistivity along the entire brick height are shown on Table 3.

| | 19-A2 | 19-B2 |
|---|---|---|
| τ (µW-PCD) [µs] | 6.8 | 4.1 |
| ρ [Ω·cm] | 1.6 | 1.1 |

*Table 3. Comparison of µW-PCD LT of bricks from ingot A2 and A2.*

The wafers from bricks 19-2A and 19-2B (2A and 2B stands for the ingot where the bricks come from) were subject to a defect engineering step. Some "as-cut" wafers were extracted from the lot and underwent a surface chemical etching in a CP4 solution and then cleaned. For surface passivation, SiNx in a plasma enhanced chemical vapor deposition (PECVD) was deposited on both surfaces. Lifetime was measured by means of quasi-steady state photo-conductance (QSSPC). Trapping was observed at low injection and, its effect corrected according to Macdonald et al., 2001. The measured initial lifetime was similar and very low (1-2 µs) for both types of wafers A and B. Then other group of wafers A and B were exposed to different phosphorous diffusion gettering (PDG) processes. The process flow is depicted in Fig. 6.

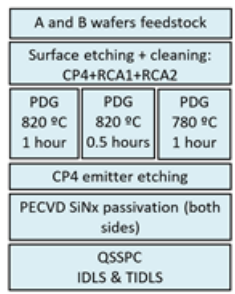

*Figure 6. Process flow for deffect engineering*

These initial lifetimes are increased one order of magnitude after a PDG step (Table 4), being the response better for the A wafers (which reach up to 80 µs) than the B ones (which in the best conditions reach 50 µs).

| τ (QSSPCD) [μs] at $10^{14}$ cm$^{-3}$ | Wafers A | Wafers B |
|---|---|---|
| PDG 780ºC, 1 h | 73 | 49 |
| PDG 820, 30 min | 51 | 25 |
| PDG 820, 1 h | 26 | 14 |

*Table 4. Results of lifetime after PDG for wafers A and B, reported at an injection level of Δn=$10^{14}$ cm$^3$ after trapping correction.*

The effectiveness of the gettering effect depends on a balance between the diffusion of impurities and their segregation to the sink layer (in this case, the P-rich layer where the impurities remain being less deleterious for the device performance). It is well known that higher temperatures promote the diffusion of impurities towards the sink layer but reduce the segregation, while at lower temperatures it is the segregation to the sink that is enhanced. A compromise can be reached, depending on the process duration and on the impurities involved (del Cañizo and Luque, 2000). In our case, the best result is obtained for the low temperature (780ºC).

Injection-level dependent lifetime curves are manipulated and linearized following the method proposed by Murphy (Murphy et al., 2012), which allows to identify two defects, that we call Defect 1 and Defect 2, whose combined contributions explain the total lifetime curves.. For each of them, the method also provides information on the ratio of electron and hole capture cross-sections (k), and the trap level in the bandgap ($E_t$), drawing a geometrical locus of their possible values. In a next step, the electron lifetime values which are compatible with the measured lifetime curves can also be drawn, giving an indication of the defect concentration (as the electron lifetime is inversely proportional to the defect concentration). I The results drawn in Fig. 7 for our UMG samples show similar k vs Et curves for wafers A and B Fig. 7, the shape of the curve ….. indicating that wafers A and B (Fig. 7(a) and Fig. 7(b)), so that it can be concluded that they are both limited by the same two defects. On the other hand, the difference in electron lifetime in Fig. 7(c) and Fig. 7(d) indicate that those defects are in higher concentration in B wafers than in A ones. .

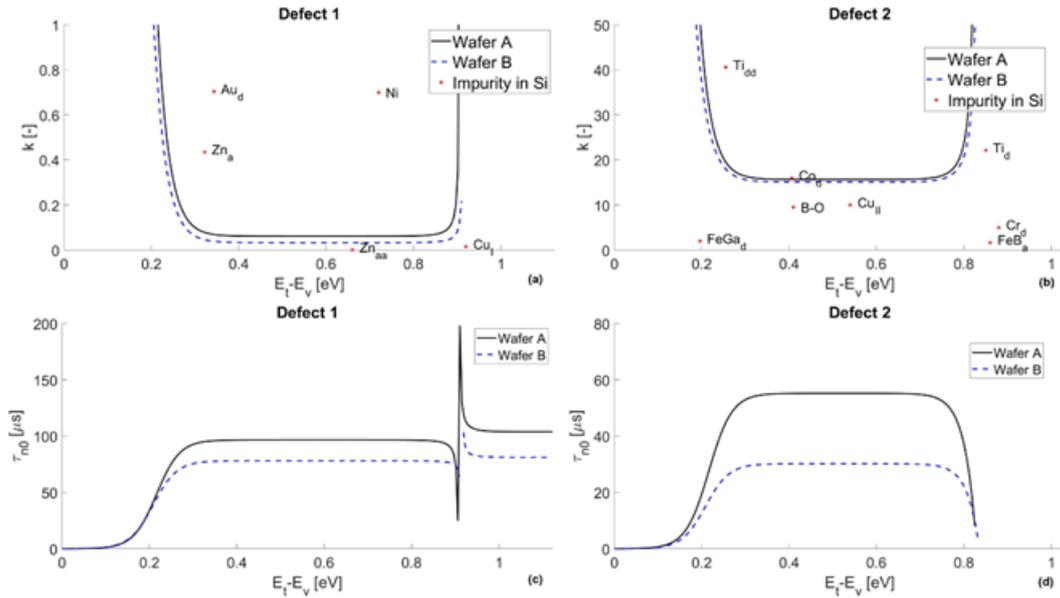

*Figure 7. Analysis of k (ratio of electron and hole capture cross-sections) as a function of the trap level in the bandgap $E_t$ (a) and (b) $\tau_{n0}$ as a function of the trap level in the bandgap $E_t$ (c) and (d) for the dominant defects in A (continuous line) and B (dashed line) wafers after the PDG. In (a) and (b) the values taken from literature for some impurities in silicon, which are close to Defect 1 and Defect 2 "fingerprints", are also shown. Note the different scales in (a) and (b).*

Fig. 7 a and b also show the values taken from literature of the ratio of capture cross-sections and trap levels of typical impurities in silicon that are close to those indicated by the geometric locus of possible values for Defect 1 and Defect 2. The results are not conclusive, though; for example, according to Fig. 7 b, it could be said that cobalt is a candidate, as its fingerprint is overlapped with its geometric locus, nevertheless, due to the chemical analysis mentioned above, the monoatomic Co can be discarded as a candidate. On the other hand, it is known that precipitated copper introduces two levels not so far from where Defect 1 and Defect 2 lie (represented as $Cu_I$ and $Cu_{II}$ in Fig. 7, according to (Macdonald et al., 2003), although the specific values are in discussion, see for instance (Inglese et al., 2016), and the presence of precipitated copper is compatible with the chemical analysis. Further characterization should be done to make any conclusive statement, keeping in mind that there exists also the possibility that these defects correspond to an impurity not well parametrized in the literature, or to an impurity that is not isolated but forms a complex with other impurities or crystal defects.

2.4. Solar cells results for type A and type B wafers:

1500 wafers of each type, from bricks 1A and 1B measured in section 2.2, were sent to a solar cell producer from the top 5 main producers in the world. Al-BSF solar cells were produced in the same production line according to the process flow depicted in Fig. 8. To dismiss the influence of any other defects different than contaminants, such as dislocations clusters, breakages, inhomogeneous emitter, etc, all the terminated cells were measured by means of electroluminescence (EL) and, in case of occurrence of the mentioned defects, rejected for this study.

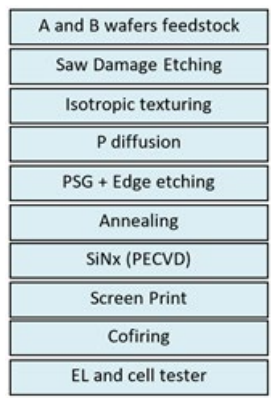

*Figure 8.* Process flow for Al-BSF cells manufacturing.

The power distribution for solar cells made of type 1A and type 1B wafers (from now named A cells and B cells respectively) is shown in Fig. 9. In the inset the average efficiency of solar cells and the lifetime measures of the bricks used to obtain the wafers A and B are presented. Type A cells show a higher average power as expected according to the lifetime measurements given in section 2.2.

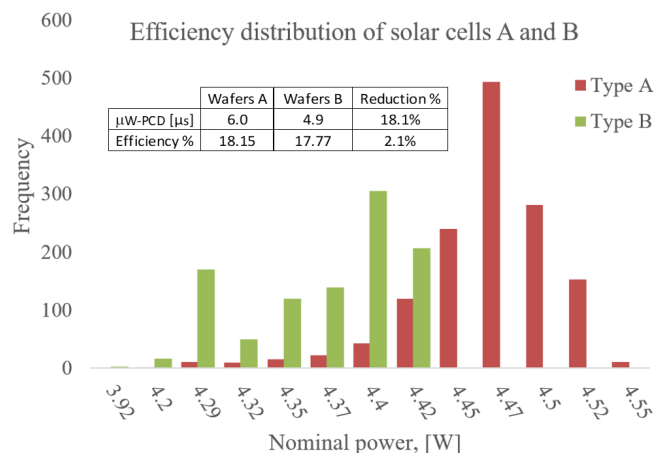

*Figure 9. Power distribution of cells A and B produced in a commercial Al-BSF production line. Inset: Comparison of LT of wafers and efficiencies of subsequent solar cells*

Besides a lower average efficiency, wafers 1B, showed a 2-peaks distribution. From this distribution, we can infer that two set of wafers, from the same ingot 1B, presented different average lifetime. Due to the mentioned production environment, the height at which the wafers were extracted was not possible to control. Thus, the extraction of the wafers from different brick heights can be the explanation, as the differences between bricks 23 and 7 were not enough to explain such a difference in cells efficiency. For example, for bricks A (19, 21), the difference in lifetime was higher, and cells A do not present a double peak distribution. The inset of Fig. 9 establishes a relation between the average lifetime reduction and the average cell efficiency loss for cells A and B.

2.5. Outdoor module degradation

After the manufacturing of solar cells, the same producer made modules using solar cells type A and solar cells made of polysilicon, both of them multicrystalline cells. To test the power degradation, 7 modules type A were installed in the same rack together with other 7 modules made of polysilicon solar cells (see Fig. 10). The location is Tudela (Spain) with an average daily irradiation of 4.7 kWh/m$^2$. The power of modules was measured by means of a wattmeter Yokogawa WT230 (basic accuracy 0.1%) and shunts of 15 A and 20 A (0.1 % accuracy). For measuring the temperature 3 PT100 TC (± 1 ºC) were placed in the backside of each module. The irradiance was obtained by measuring the current generated by the reference module of the same technology. The daily energy production along 24 months of Sun exposure was recorded and the performance ratio at 25ºC calculated according to Eq. 1:

$$PR_{25} = \frac{\dfrac{E_{PV,d}\,[Wh]}{(1 + \gamma \cdot (T_{PV,d} - 25))}}{\dfrac{Pn\,[W]}{1000\,[\frac{W}{m^2}]} \cdot Gs\,[\frac{Wh}{m^2}]}$$

Equation 1

Where $E_{PV,d}$ is the daily energy produced by the string of modules, $\gamma$ is the power temperature coefficient of the module, $T_{PV,d}$ is the average temperature of the module along the day, $P_n$ is the nominal power of the string of modules and $G_s$ the daily solar irradiation measured by the reference module.

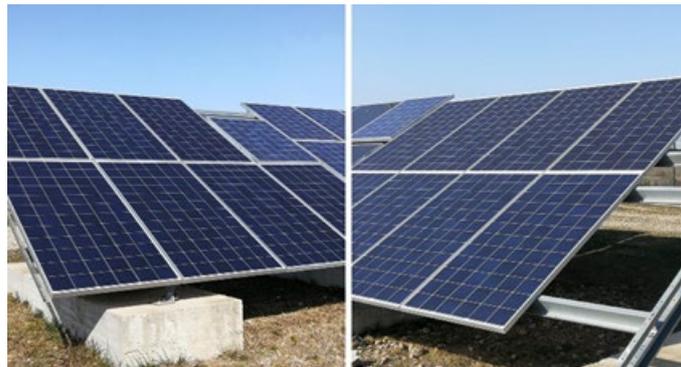

*Figure 10. Modules installation. Left: modules made of A wafers. Right: modules made of standard polysilicon wafers. Both modules are multicrystalline.*

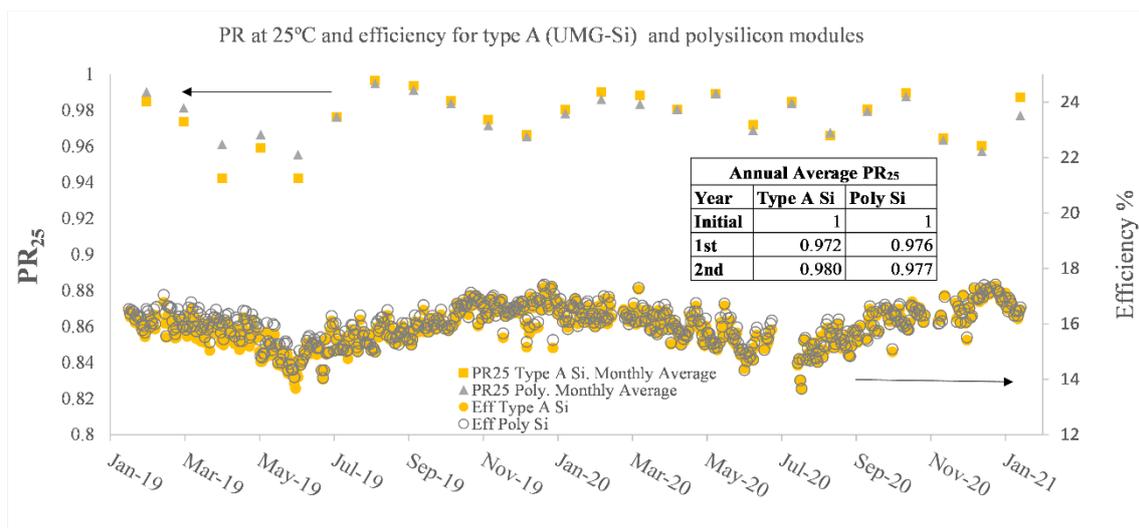

*Figure 11. Efficiency and Monthly average of PR calculated at 25ºC for UMG Si (type A) and polysilicon. 24 months of degradation. Absent data correspond to interventions on the equipment.*

As mentioned in the introduction, the authors want to detect any early degradation that could affect UMG silicon modules more than polysilicon. For that purpose, module efficiency and monthly average of $PR_{25}$ (Ishii and Masuda, 2017) are shown in Fig. 11. This figure clearly shows that power degradation for both types of modules (UMG-Si and Poly) are equivalent confirming other works publications carried out on similar silicon materials (Huang et al., 2016; Sánchez et al., 2018). In the inset of Fig.11, we can see the annual average of $PR_{25}$ for the two years of degradation. The first year, the $PR_{25}$ has decreased from 1 to 0.972 por type A UMG-Si and to 0.976 for poly Si, which is in concordance with the initial light induced degradation (LID) of Al-BSF solar cells (Pingel et al., 2010). After 2 years of degradation, the $PR_{25}$ is even higher for type A modules ($PR_{25}$ = 0.980) than for polysilicon modules ($PR_{25}$ = 0.977),

2.6. Life Cycle Assessment comparison of FerroSolar UMG silicon and Hydrochlorination Siemens route polysilicon.

Besides the technical and performance aspects, the climate change (CC) impact category of Ferrosolar´s UMG silicon compared to that related to polysilicon coming from the "gas route" has been also studied. CC refers to the global emissions of GHG, expressed in equivalent kg of carbon dioxide (kg $CO_2$ eq) for each kilogram of manufactured silicon. A life cycle assessment (LCA) approach has been used following the methodology standards [(ISO, 2006a, 2006b)] and the PVPS Methodology Guidelines for PV electricity (Frischknecht et al., 2016). The results presented here are part of a more complete LCA and the details can be consulted elsewhere (Méndez et al., 2021), that comprehends all the environmental impact categories of all the steps in the multicrystalline silicon PV value chain, from quartz mining to end-of-life PV installation, for both UMG and conventional polysilicon. Life-cycle inventory tables, where the whole set of data for analysis is collected, have been reviewed and adapted to the current industrial status. This way, all the processes, materials used, emissions, etc. are the most updated, up to the authors knowledge.
As the process of UMG production assessed in this study takes place in Spain (Puertollano, (Forniés et al., 2019)), the Spanish electricity mix has been used for calculations. Nevertheless, most of the PV materials, including the solar grade silicon, are produced in China, hence, results using Chinese electricity mix as a primary energy input are also presented. Fig. 12 shows the electricity mix of China (BP, 2018) and Spain (Red Eléctrica de España, 2019) in 2019, considered in this study. Main reference data have been adapted from ecoinvent database v3.5.

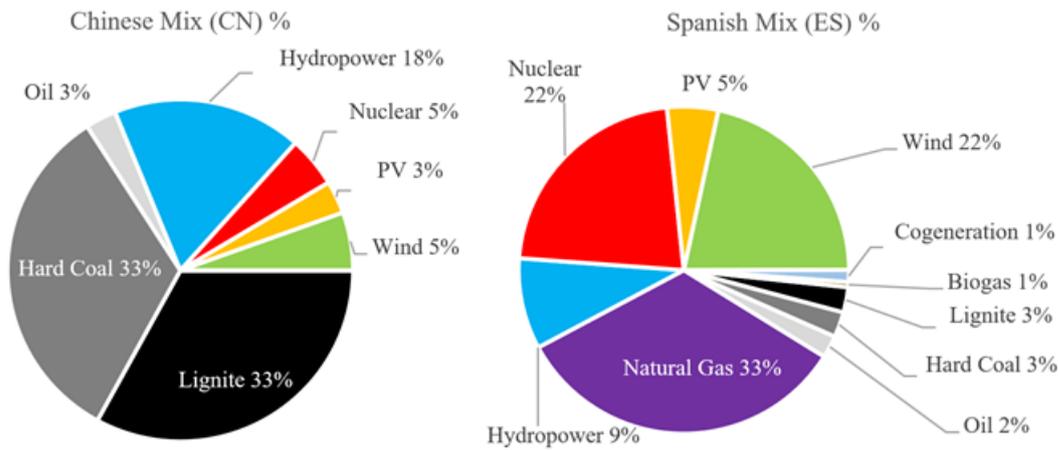

Figure 12. Chinese and Spanish electricity mix.

This silicon is to be compared with the standard polysilicon used for the PV industry. For a more updated analysis, hydrochlorination is considered, for being the state-of-the-art process for obtaining polysilicon. It is based on the obtention of purified silicon by means of distillation of trichlorosilane (TCS) and a subsequent CVD deposition in a Siemens reactor. The particularity of this process is that it takes the advantage of using a by-product, silicon tetrachloride (STC), recycling it in the beginning of the process, to transform the metallurgical grade silicon into TCS through the following reaction (Forniés et al., 2016):

$3SiCl_4(g) + Si(s) + 2 H_2(g) \rightarrow 4SiHCl_3(g)$

Moreover, the recent improvements in energy saving and increase of productivity implemented in the Siemens process have been considered (Shravan et al, 2017; Woodhouse et al., 2019).

Fig. 13 shows the results for UMG-Si versus polysilicon with both Chinese and Spanish electricity mix. UMG is nearly three times lower in GHG emissions when compare to the conventional polysilicon. When different electricity mix is considered, the Spanish shows a minor impact due to its more renewable energy contribution to the mix.

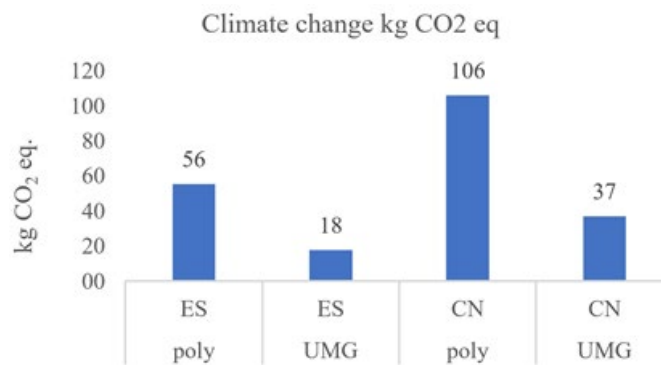

Figure 13. kg of equivalent $CO_2$ for each kg of silicon produced. Two different energy mixes has been used to make the calcualtions, spanish mix (ES) and chinese mix (CN).

Other studies, comparing UMG and polysilicon, have been made before by de Wild-Scholten (de Wild-Scholten et al., 2008). In this case Norwegian and UCTE (Union for the Co-ordination of Transmission of Electricity) electricity mixes are considered. These results are in the same range than present work, although different assumptions and databases were considered. More recent research related to UMG silicon production environmental impact has been carried out in China (Yu et al., 2017). Although the aggregated manner in with the results are presented makes impossible to carry out a detailed study, their results for China follow the same trend as the ones presented in this work, being UMG environmental impact significantly lower than polysilicon. This is what actually is expected, as the main contribution, especially to the CC category, in both processes, comes from the important amount of electricity consumed, which is significantly lower in the case of UMG: 25 kWh/kg vs. 85 kWh/kg in the case of polysilicon (Shravan K. Chunduri, 2017; Xie et al., 2018).

**Conclusions:**

A complementary study of UMG silicon to that made in other studies (Forniés et al., 2019, 2018) is presented here. Chemical analysis at the wafer level and defect study was applied here although. Whilst the dopant concentration has been determined by means of IPC-MS, the same technique could not give conclusive results about metals concentration different than the evidence that Co, Ni and Cr cannot be considered as the cause of lifetime reduction for wafers B, at least as monoatomic defects. Regarding the defect analysis, by applying the injection dependent lifetime spectroscopy (IDLS) it has been shown that A and B wafers are limited by the same two defects, being their concentration higher in the B wafers. The comparison with the fingerprints of metal impurities dissolved in silicon helps to discard many of them as limiting defects in UMG silicon; on the other hand, precipitated copper shows two levels that are compatible with the IDLS results, making it a sound candidate for being the limiting defect, althoughfurther research should be done to confirm this point. At the cell level, we can stablish a clear relationship between the measured lifetime at the ingot level and the efficiency obtained at the cell level for both types of wafer (A and B). While the loss of efficiency due to the usage of UMG had been clearly determined and quantified in the previous studies mentioned above, the degradation had not been determined so far. This is another contribution of present work, where the outdoor degradation clearly shows the same degradation of modules made of polysilicon and UMG silicon. To finalize the study of UMG silicon, a life cycle assessment has been carried out showing the considerably lower carbon footprint of UMG silicon compared with polysilicon.

**Acknowledgements:**

The Centro para el Desarrollo Tecnológico Industrial (CDTI) and the Spanish Agencia Estatal de Investigación are acknowledged for partial funding through the project "Low Cost High Efficient and Reliable UMG PV cells (CHEER-UP)", as part of the SOLAR-ERA.NET Cofund 2 Call. Mallory Jensen, former member of MIT PV lab, now at Sunpower, is acknowledged for continuous support and advice about the IDLS analysis. Acciona (www.acciona.com)  and Universidad Pública de Navarra (UPNA) are acknowledged for their set up and measurements of outdoor module degradation.